\documentclass[prb,twocolumn,showpacs,showkeys,preprintnumbers,amsmath,amssymb,superscriptaddress,aps,10pt]{revtex4-1}
\usepackage{graphicx}
\usepackage{dcolumn}
\usepackage{bm}
\usepackage{natbib}
\usepackage{amsmath}
\usepackage{color}
\usepackage{ulem}

\begin{document}
%
\title[PSL on NbSe2]{Phase slip lines in superconducting few-layer NbSe$_2$ crystals}
\author{Nicola Paradiso}\email{nicola.paradiso@physik.uni-regensburg.de}
\author{Anh-Tuan Nguyen}
\author{Karl Enzo Kloss}
\author{Christoph Strunk}
\affiliation{Institut f\"ur Experimentelle und Angewandte Physik, University of Regensburg}
%
%
%
\begin{abstract}	
	
We show the results of two-terminal and four-terminal transport measurements on few-layer NbSe$_2$ devices at large current bias. 
In all the samples measured, transport characteristics at high bias are dominated by a series of resistance jumps due to nucleation of phase slip lines, the two dimensional analogue of phase slip centers. In point contact devices the relatively simple and homogeneous geometry enables a quantitative comparison with the model of Skocpol, Beasley and Tinkham. 
In extended crystals the nucleation of a single phase slip line can be induced by mechanical stress of a region whose width is comparable to the charge imbalance equilibration length.

\end{abstract}
%
\maketitle

At sufficiently large current bias any superconductor (SC) shows dissipation. The ultimate limit is the pair-breaking current described by the Ginzburg-Landau theory.~\cite{Ginzburg1958,BardeenRPM1962} Before that limit, dissipation may emerge as a result of the motion of Abrikosov vortices, which can be introduced by an external field or by the sample current itself. However, in the late '60s experiments on tin whiskers~\cite{Webb1968} showed that, even in absence of vortices, dissipation is observed before the pair-breaking current density is reached owing to the nucleation of phase slip centers (PSCs). 

PSCs were first observed in 1D superconducting systems, ~where the transverse size is much smaller than both the coherence length $\xi_{GL}$ and the magnetic penetration depth $\lambda$. Accordingly, the models used to describe them were all inherently one dimensional.~\cite{SBT1974,Tinkham1979,Ivlev1984} It was soon found, however, that 2D systems can support an analogue of PSCs as well, the so called phase slip lines (PSLs).~\cite{Ogushi1972,Volotskaya1981,Dmitrenko1996} The physical mechanism of the PSLs is markedly different from that of PSCs. It was found from numerical simulations based on time-dependent Ginzburg Landau (TDGL) equations~\cite{Weber1991,Andronov1993} that PSLs are a sort of rivers of fast vortices and antivortices (called \textit{kinematic vortices}~\cite{Andronov1993}) that annihilate in the middle of the sample.~\cite{Berdiyorov2009,Berdiyorov2014} Despite the physical difference, PSLs produce similar current-voltage characteristics (IVC) as PSCs, namely, discontinuous voltage jumps separating linear portions whose extrapolation on the current axis leads to a finite excess current.~\cite{Dmitrenko1996,Sivakov2003,Dmitriev2005,Falk2007,ZhouPRB2007}

In ordinary polycrystalline or amorphous superconducting films, it is not easy to observe signatures of PSLs. The main experimental difficulty~\cite{Dmitriev2005} is to prevent the formation of Abrikosov vortices by keeping a sufficiently high energetic barrier at the edges.~\cite{Shmidt1970} Moreover, it is crucial to ensure a good heat dissipation since at the large current densities close to the PSL nucleation Joule heating  can easily smear the IVC.~\cite{Dmitrenko1996}

All these limitations are absent in van der Waals SCs.~\cite{SaitoReview2016,NovoselovReview2016} The nowadays standard exfoliation techniques make it possible to obtain clean  monocrystalline devices consisting of one or few atomic layers, thus with a large Pearl screening length $\lambda_{\perp}=\lambda^2/d$. The edge roughness is on the nanometer scale, i.e.~much sharper than the typical roughness of devices patterned by electron beam lithography (EBL). Encapsulation in hBN, besides preventing oxidation and contamination, provides an efficient thermal sink.~\cite{Zhou2014,Liu2017} Therefore, one would expect that PSLs should dominate the IVC of van der Waals SC for current values well below the pair-breaking value.

The most studied van der Waals SC is the 2H polytype of NbSe$_2$ (in the following simply indicated as NbSe$_2$).~\cite{Staley2009,ElBana2013,Cao2015,Tsen2015,Xi2015Ising,Xi2015CDW,Xi2016gate,Yabuki2016,Xing2017,Wang2017,Nguyen2017,Khestanova2018,Dvir2018,Lian2018,delaBarrera2018} This material can be easily exfoliated in few-layer-thick flakes and then encapsulated in hBN.~\cite{Cao2015} Typical devices for transport measurements have a size of several micrometers. Several key results have been obtained in the recent years on this material. It was shown~\cite{Cao2015,Xi2015CDW} that both the superconducting and the charge-density-wave phase survive even in monolayer crystals and that Ising superconductivity is observed owing to the large spin-orbit pseudo-magnetic field.~\cite{Xi2015Ising,Xing2017} 
However, such experiments have focused on the low bias regime. To date, very little is known about the high current bias regime. We expect that the 2D character of exfoliated NbSe$_2$ crystals allows for the observation of PSLs possibly even below the critical temperature.

%
\begin{figure*}[tb]
\includegraphics[width=2\columnwidth]{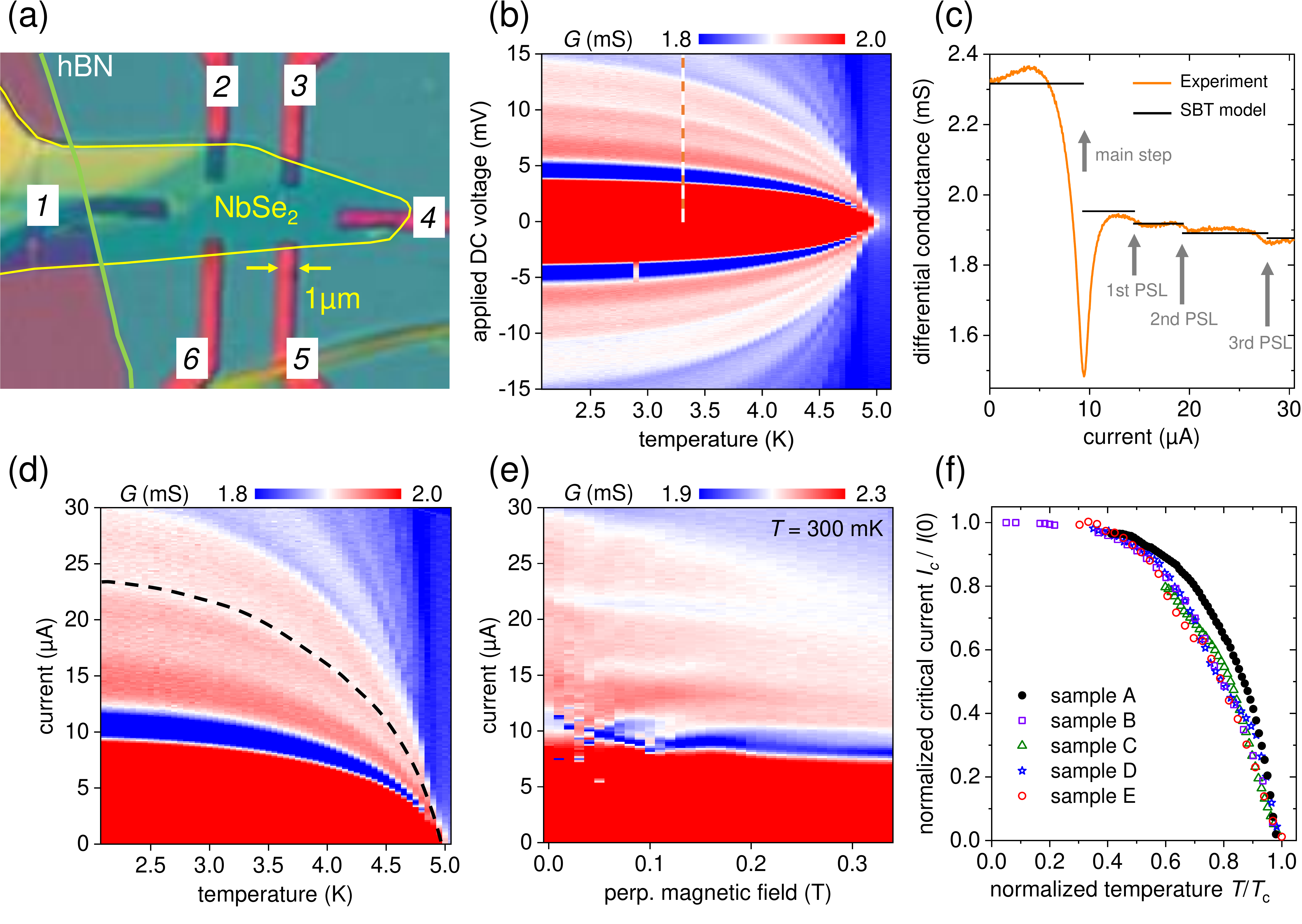}
\caption{(a) Optical picture of the first sample discussed in this Letter (sample $A$). A trilayer flake of NbSe$_2$ is stamped on prepatterned Au contacts. The crystal shows several terraces of different thickness. The crystal portion overlapping contact \textit{4}, \textit{5} and \textit{6} is three layer-thick. (b) Color plot of the differential conductance between contact \textit{1} and \textit{5} as a function of voltage bias and temperature. (c) Measured differential conductance between the same contacts, plotted as a function of current for $T=3.3$~K (orange line) together with the values calculated using the Skocpol-Beasley-Tinkham model with the same parameters as in Fig.~\ref{fig:simulcolplot}(b) (see text). The bias range corresponds to the orange-white dashed line in panel (b). (d) Differential conductance plotted as a function of \textit{current} (for positive bias) and temperature. By increasing the bias, the conductance decreases in steps. The edge of one of such steps is highlighted with a dashed line. (e) Differential conductance as a function of perpendicular magnetic field at $T=300$~mK. This measurement was performed in a subsequent cool-down compared to the one reported in the previous panels. (f) Critical current as a function of temperature for several few-layer NbSe$_2$ devices measured in a four-terminal configuration (empty symbols). For each sample the abscissas and the ordinates of the data points have been normalized to the critical temperature and to the extrapolated critical current at 0~K, respectively. The black dots refer to the edge of the conductance step highlighted in panel (d). 
}
\label{fig:expcolplot1}
\end{figure*}
In this Letter we present the results of finite-bias measurements on devices based on few-layer (bi- or trilayer) NbSe$_2$ crystals. We shall first discuss two-terminal measurements on a point contact between Au electrode and trilayer NbSe$_2$. Such measurements allow us to quantitatively understand the high bias transport features in terms of sequential nucleation of PSLs. In the second part, we discuss four-terminal measurements on extended flakes, where clear evidences of PSL nucleation have been observed in all devices measured so far. Finally, we show that it is possible to create an artificial nucleation site for PSLs by mechanically stressing a NbSe$_2$ flake.

The first sample here described (sample $A$) was fabricated starting from a doped Si substrate capped with a 285~nm-thick thermal SiO$_2$ layer. Six ohmic contacts were defined by EBL followed by thermal evaporation of a 17~nm-thick Ti/Au film (2~nm Ti as sticking layer followed by 15~nm Au as top layer). Residuals of resist were removed by a plasma oxygen treatment. A trilayer NbSe$_2$ flake (thickness $d=1.8$~nm) was then transferred onto the contacts using a dry-transfer technique.~\cite{Castellanos2014} NbSe$_2$ commercial crystals (HQgraphene) were exfoliated onto scotch tape and then transferred onto a polydimethylsiloxane (PDMS) film. The latter was then stamped on the prepatterned electrodes. Accurate control of the alignment during stamping was obtained using a micromanipulator.
Finally, a several layer-thick flake of hBN was stamped on the NbSe$_2$ flake using the same technique. The hBN layer is needed to protect the NbSe$_2$ crystal underneath from oxidation and to improve the thermal coupling with the substrate. 

\begin{figure*}[tb]
\includegraphics[width=2\columnwidth]{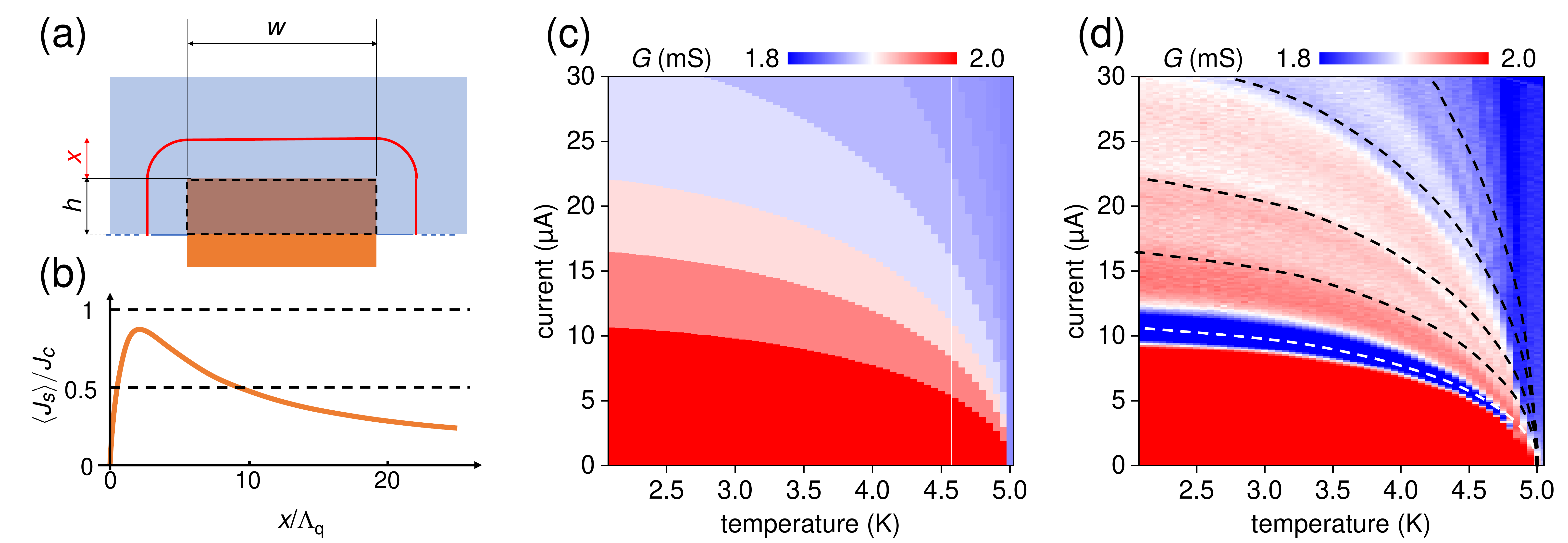}
\caption{(a) Sketch of the region near the contact \textit{5} of sample $A$. The dashed rectangle indicates the overlap region between (proximitized) NbSe$_2$ and Au electrode. The red curve indicates a line of constant current density. In sample $A$, $w\approx 1$~$\mu$m and $h\approx 0.2$~$\mu$m. (b) Schematic behavior of the time-averaged supercurrent density as a function of the distance from the proximitized region $x$ expressed in units of the charge imbalance equilibration length $\Lambda_q$. For this graph we assumed $L\equiv (w+2h)/\pi=5\Lambda_q$. (c) Differential conductance calculated using the Skocpol-Beasley-Tinkham model with the parameters discussed in the text.  (d) Color plot of the measured differential conductance plotted as a function of current and temperature. The overlay lines correspond to the step edges in the previous panel.
}
\label{fig:simulcolplot}
\end{figure*}

An optical image of sample $A$ is shown in Fig.~\ref{fig:expcolplot1}(a). The NbSe$_2$ flake consists of three areas: the thinnest (three layer-thick) area overlaps three contacts, namely the two bottom contacts (\textit{5} and \textit{6}) and the one on the right 
(\textit{4}); the medium-thick (10 layers) area overlaps the two contacts on top (\textit{2} and \textit{3}); finally, a large and bulky region adheres on the left contact (\textit{1}) over a macroscopic area. Owing to its low resistance, this latter contact  is used as a source in all two-terminal measurements. As a drain contact, we used one of the other contacts. All but the bottom left contact show a very low contact resistance in the few hundred ohm range at low temperatures. In contrast, contact \textit{6}, possibly due to resist residuals on the electrode surface, behaves like an opaque tunnel junction whose finite-bias differential conductance reflects the NbSe$_2$ density of states (see Supporting Information). We focused on two-terminal measurements at finite bias between contacts \textit{1} and \textit{5}. The overlap area between NbSe$_2$ and the Au electrode of contact \textit{5} is only about 0.1--0.2~$\mu$m$^2$. Being so small makes this Au-NbSe$_2$ interface much more homogeneous than the others. Hence, transport data are quantitatively easier to interpret, although qualitatively similar to what observed for the other contacts of this sample and for other samples as well (details on these additional measurements are reported in the Supporting Information). 

Transport measurements have been performed in two different cryostats with base temperature of 2~K and 300~mK, respectively. The latter one is equipped with a superconducting coil for measurements in finite magnetic field. Two-terminal differential conductance is measured by lock-in technique. A small AC excitation  $\delta V_{AC}=20$~$\mu$V is superimposed to the swept DC voltage bias $V_{DC}$. Differential (AC) and direct current signal are then simultaneously recorded.

The color plot in Fig.~\ref{fig:expcolplot1}(b) shows the differential conductance measured between contacts \textit{1}--\textit{5} as a function of finite voltage bias and temperature. Fig.~\ref{fig:expcolplot1}(c) displays the differential conductance as a function of current for positive bias and for $T=3.3$~K. This corresponds to the orange-white dashed segment in panel (b). We can identify three main features:  a slow parabolic increase of the conductance at relatively low bias is followed by a sharp dip and then by a series of downward steps. The first step is much larger than the others.
The spacing between the edges of the small steps increases with the bias while their amplitude slightly decreases. It is important to notice that all the features here observed occur at voltage bias $V_{DC}$ much larger than the superconducting gap of NbSe$_2$, which is of the order of $1.764k_BT_c\approx 0.76$~meV for $T_c=5$~K. For $V_{DC}$ of the order of the gap, the conductance trace is featureless. 
The latter observation indicates that the observed features are not related to any quasi-particle tunneling process.

A clear clue on the origin of the steps comes from the observation of the temperature dependence of the step edges. We have highlighted one of them in  Fig.~\ref{fig:expcolplot1}(d) (dashed line), which shows the differential conductance as a function of \textit{current} and temperature for positive bias.   All step edges follow the same dependence as a function of temperature. 
It is instructive to compare the current at the step edges to the overall critical current for the onset of dissipation (the latter defined as the current at which the voltage drop exceeds 1~$\mu$V, see Supporting Information for further details) in  few-layer NbSe$_2$ devices. 
Figure~\ref{fig:expcolplot1}(f) shows (empty symbols) the value of the critical current for several few-layer NbSe$_2$ flakes measured in four-terminal configuration. For each sample, the current and temperature values have been rescaled to keep into account the individual  values of $T_c$ and the different critical current values at low temperature. All these curves have the same temperature dependence, which is similar to that of the step edges in sample $A$ (black dots). Interestingly, except deviations near $T_c$, the temperature dependence of all the curves seems to reproduce the temperature dependence of the BCS gap (see discussion in the Supporting Information). 
This fact is non-trivial since, e.g., for dirty SCs the depairing critical current density  far from $T_c$ is expected to scale as~\cite{BardeenRPM1962,Romijn1982} $[1-(T/T_c)^2]^{3/2}$. 
What is important for the present discussion is the fact that the temperature dependence of the critical current in plain NbSe$_2$ devices is the same as that of the step edges in  Fig.~\ref{fig:expcolplot1}(d). This fact, together with the peculiar staircase pattern, indicates that conductance steps are a signature of the sequential PSL nucleation near the contact interface.

A further demonstration of the PSL nature of the steps is given by conductance measurement at finite perpendicular magnetic field. As mentioned above, the introduction of Abrikosov vortices rapidly disrupts the PSL-induced steps in the IVC. Numerical simulations based on TDGL equations~\cite{Berdiyorov2009} confirm that, already at fields of the order of 10\% of the critical value, the step structure in the IVC is almost completely washed out. Figure~\ref{fig:expcolplot1}(e) shows a color plot of the differential conductance as a function of current and perpendicular magnetic field measured at $T=300$~mK. We notice that, except for the first (and largest) step, all the other steps are washed out already at 300~mT, which is only 10\% of the critical field. On the one hand, this confirms that all the steps except the first are signatures of PSLs. On the other hand, the persistence of the first step at high magnetic fields indicates that its nature is different.

The above evidences allow us to construct a quantitative description of the high bias behavior of the point contact. A sketch of the region around contact \textit{5} is shown in Fig.~\ref{fig:simulcolplot}(a).  The low tunnel resistance of the point contact implies that the overlap region between Au and NbSe$_2$ (dashed rectangle in the sketch) must be at least partially proximitized, since a low transparency contact would lead to a much lower subgap conductance. Therefore, the critical current in this region is lower than that in the rest of the crystal. By gradually increasing the current, this region will then turn normal first. This corresponds to the first step in the conductance plot. The critical current $I_p$ corresponding to this step is suppressed for perpendicular magnetic fields of the order of the critical field $H_{c2}\approx 3$~T.~\cite{Cao2015}  The first step is therefore only weakly affected by moderate magnetic fields, as observed in Fig.~\ref{fig:expcolplot1}(e).

When the overlap region turns normal, the time-averaged supercurrent \textit{density} $\langle J_s\rangle$ depends non-monotonically on the distance $x$ from the normal-superconducting (NS) interface, as sketched in Fig.~\ref{fig:simulcolplot}(b).
At the NS interface ($x=0$) the supercurrent is zero: all of the current is carried by unpaired electrons. The conversion of normal current into supercurrent at a NS interface is mediated by inelastic scattering processes that equilibrate the local charge imbalance.~\cite{SBT1974,Pippard1971,Tinkhambook} As a result, the normal current density decreases nearly exponentially as a function of $x$ with a decay length scale $\Lambda_q$, the charge imbalance equilibration length. Correspondingly, $\langle J_s\rangle$ increases  and within a distance of the order of $\Lambda_q$ almost all the current is carried as supercurrent. At larger distances the total current (and thus the supercurrent) density decreases because the cross section of the SC increases. Let us consider a line connecting points at the same distance $x$ from the NS interface (red line in Fig.~\ref{fig:simulcolplot}(a)). The cross section of the SC corresponding to this line is the SC thickness $d$ times the length of the line $w+2h+\pi x$, where $w\approx 1$~$\mu$m and $h\approx 0.2$~$\mu$m are the width and height of the proximitized region, respectively (see Fig.~\ref{fig:simulcolplot}(a)). Hence, $\langle J_s\rangle$ as a function of $x$ scales approximately as
\begin{equation}
\langle J_s(x) \rangle \propto \frac{1-e^{-x/\Lambda_q}}{w+2h+\pi x},
\label{eq:timeaveragedJs}
\end{equation}
where the numerator describes the normal-to-supercurrent conversion near $x=0$ and the denominator accounts for the increase of the cross section at large $x$. This behavior is sketched in Fig.~\ref{fig:simulcolplot}(b):  $\langle J_s\rangle$ as a function of $x$ looks like a distorted bell with a maximum near  $x=\Lambda_q$ and a tail that scales as $(w+2h+\pi x)^{-1}$. The width of the bell is approximately $L\equiv (w+2h)/\pi\approx 445$~nm, where $L$ is defined as the value of $x$ such that the length $(w+2h+\pi x)$ is twice that for $x=0$. 

By increasing the current bias, a PSL will nucleate when  $\langle J_s(x)\rangle$ locally exceeds the critical current density (see Supporting Information for a more detailed illustration of the process). By further increasing the current, more PSLs will sequentially nucleate. They will mostly concentrate within a region of size $L$, since  $\langle J_s(x)\rangle$ drops very quickly for large $x$. The convergence of the current lines near the narrow NS interface tends to confine the PSLs in a region whose size is comparable to the contact size. This allows us to adapt the analytical model of Skocpol, Beasley and Tinkham~\cite{SBT1974} (SBT) for the nucleation of PSCs in a homogeneous wire of length $L$. 

The differential resistance as a function of current for a system of $n$ interacting PSLs confined in a length $L$ is given by~\cite{Tinkham1979} 
\begin{equation}
R^{(n)}_{PSL}=R_0\frac{2\Lambda_qn}{L}\tanh\left(\frac{L}{2\Lambda_qn}\right),
\label{eq:respsl}  
\end{equation}
where $R_0=\rho L/A$, with $\rho$ the low temperature limit of the resistivity in the normal state and $A$ the transverse section of the crystal, which is approximately $A\approx d(w+2h)=d\pi L$. Therefore $R_0=\rho/(d\pi)\equiv R_{\square}/\pi$, where $R_{\square}$ is the sheet resistance of trilayer NbSe$_2$, which is expected to be~\cite{Xi2015Ising,Tsen2015,Soto2007} approximately 66~$\Omega$.

The nucleation of the $n$-th PSL occurs when the current equals (see Eq.~3 in Ref.~\onlinecite{Tinkham1979})
\begin{equation}
I_n(T)=I_{c}(T)\frac{\cosh\left[\frac{L}{2\Lambda_qn}\right] - b}{\cosh\left[\frac{L}{2\Lambda_qn}\right]- 1},
\label{eq:biasstep}  
\end{equation}
where $b\approx 0.5$ is ratio between time-averaged and maximum supercurrent,~\cite{Tinkham1979,Tinkhambook} and $I_c$ is the critical current of the SC.

Finally, we need to take into account a series resistance $R_b$ due to the finite interface transparency plus the resistance of the cryostat cables. Also, for current higher than the critical value $I_{p}$, we add the normal resistance of the proximitized region $R_{p}$.
The differential conductance as a function of current and temperature is thus given by
\begin{equation}
G(I,T)=\left(R_b+R_p\Theta(I-I_p(T))+R^{(n)}_{PSL}(I,T) \right)^{-1},
\label{eq:Gtot}
\end{equation}
where the current dependence of $R^{(n)}_{PSL}$ is given by Eq.~\ref{eq:biasstep} and the temperature dependence of the critical current values $I_c(T)$ and $I_p(T)$ is simply proportional to that of the BCS gap $\Delta(T)$, as a result of the empirical observation discussed above. 
The resulting two-terminal differential conductance is plotted in Fig.~\ref{fig:simulcolplot}(c). 
The input parameters $R_b=431$~$\Omega$, $R_p=80.2$~$\Omega$, $I_c(0)=16$~$\mu$A, $I_p(0)=11$~$\mu$A and $T_c=4.95$~K can be immediately obtained from the experimental data. The last two parameters, i.e.~$2\Lambda_q/L=0.355$ and $R_{\square}=84$~$\Omega$ are chosen in order to fit the distribution and the amplitude of the PSL steps. In Fig.~\ref{fig:simulcolplot}(d) we highlight the conductance step edges as found in the model calculation (dashed lines) on top of the experimental data. We observe that, despite the drastic approximations behind our simple model, the agreement is satisfactory, especially for low number of PSLs. As the number of PSLs increases, the assumption that PSLs are rigidly confined in a length $L$ becomes more and more inaccurate. This produces the deviations observed in Fig.~\ref{fig:simulcolplot}(d) between the experimental and the calculated position of the conductance step edges for high number of PSLs. 

The ratio $2\Lambda_q/L$ obtained from the fit allows us to estimate the charge imbalance equilibration length $\Lambda_q\approx 0.355L/2=80$~nm. Assuming a mean free path~\cite{Tsen2015,Dvir2018} of about 30~nm, a Fermi velocity of about~\cite{Kiss2007} 10$^5$~m/s and a diffusion length~\cite{Dvir2018} $D=v_F\ell/3=10^{-3}$~m$^2$/s one can deduce a charge imbalance relaxation time $\tau_q=\Lambda_q^2/D\approx 6$~ps.

\begin{figure}[tb]
\includegraphics[width=\columnwidth]{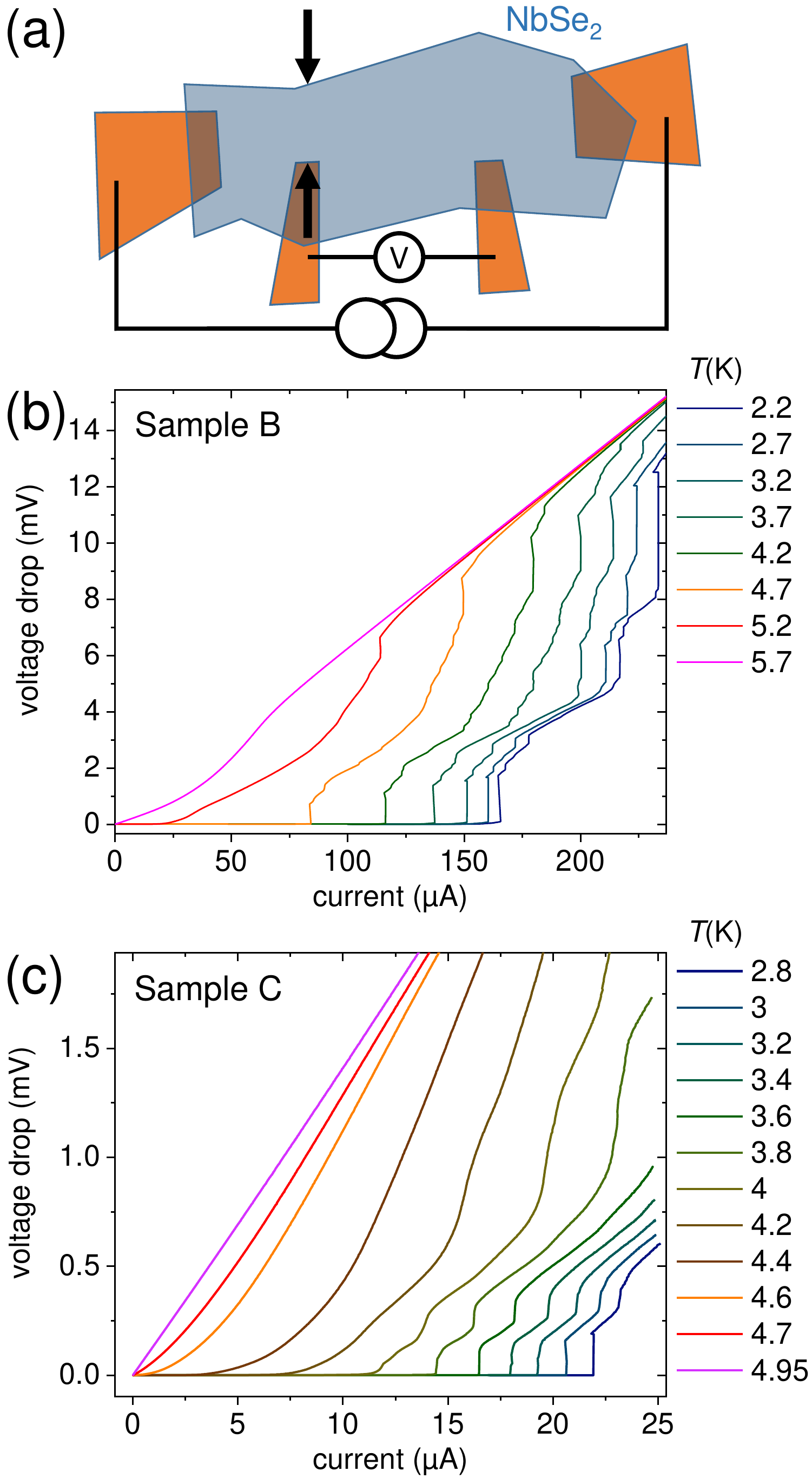}
\caption{(a) Scheme for the four-terminal measurements discussed in the text. The arrows indicate where the PSLs would preferentially nucleate if at high current bias the regions underneath the voltage probe contacts turn normal. (b) Four-terminal current-voltage characteristics measured for different temperatures in sample $B$ (trilayer NbSe$_2$).  (c) The same for sample $C$ (bilayer NbSe$_2$). 
}
\label{fig:4term}
\end{figure}

Having found that transport in the vicinity of point contacts is dominated by the nucleation of PSLs, it is natural to expect that PSLs should also occur in  four-terminal measurements on extended flakes.
Figure~\ref{fig:4term}(a) sketches the geometry of a typical four-terminal measurement. The contact electrodes can be either prepatterned Au strips or many-layer graphite flakes stamped on top of NbSe$_2$. The graphs in Fig.~\ref{fig:4term}(b,c) show IVC of two selected samples for different temperatures (more data is shown in the Supporting Information). In \textit{all} the samples we measured so far, we have observed signatures of PSL nucleation: above a threshold current the IVC shows voltage jumps with distinct linear segments. The extrapolation point on the current axis of such linear segments reveals a finite and large excess current. While the qualitative behavior of all the measured samples is similar, the details of the IVC are sample-dependent, as e.g.~the position and amplitude of the steps. Such variance in the IVC is closely related to the diverse geometry of the NbSe$_2$ flakes obtained by exfoliation. Different geometries clearly lead to different PSL distributions and thus to different step patterns in the IVC. In addition, the effective geometry of the supercurrent flow also depends on the position of the contacts.
From the point contact measurement discussed above, we learned that, for an ultra-thin SC, normal contacts suppress the critical current density owing to the proximity effect. In devices based on exfoliated NbSe$_2$ the contacts occupy a substantial fraction of the flake area,~\cite{Cao2015,Tsen2015,Xi2015Ising} rather than being point-like. Therefore, voltage probes distort the supercurrent path when, at high current bias, the NbSe$_2$ region underneath is forced normal conducting. For example, in the sketch of  Fig.~\ref{fig:4term}(a) if the current bias is large enough to turn the contact regions  normal, then the supercurrent will concentrate in the region between the black arrows. This will clearly be a preferred nucleation site for PSLs.

Another factor relevant for the PSL arrangement is the different distribution of defects in each exfoliated flake. In clean van der Waals materials such defects might be due to bubbles of air, water or  hydrocarbon contaminants~\cite{Haigh2012} trapped underneath the flake during the stamping process. Such bubbles might stress the crystal and affect the arrangement of PSLs.

\begin{figure}[tb]
\includegraphics[width=\columnwidth]{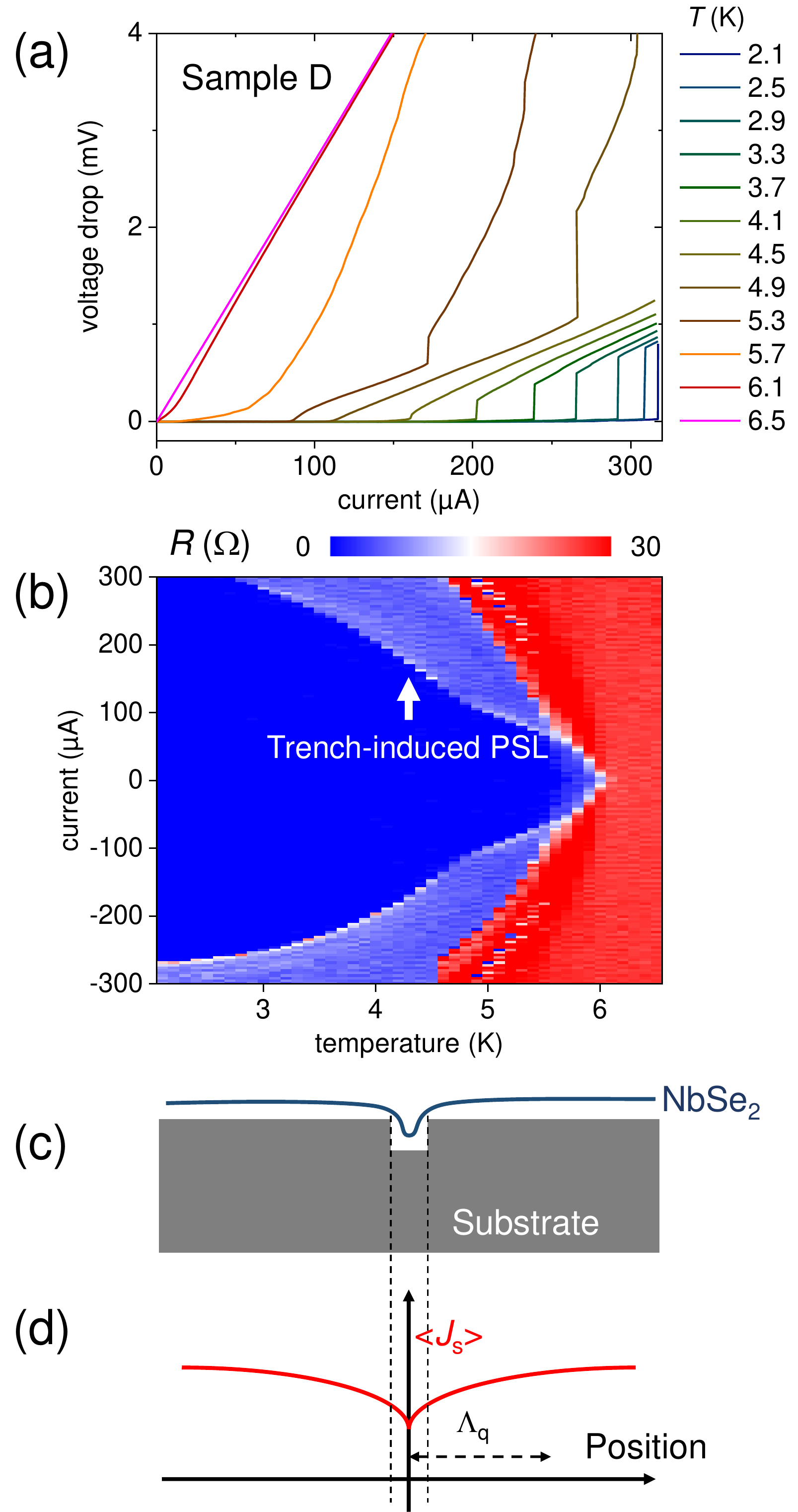}
\caption{(a) Current-voltage characteristics measured for different temperatures in sample $D$ (trilayer NbSe$_2$). (b) Color plot of the differential resistance as a function of current bias and temperature. (c) Sketch of the central region of sample $D$ (side view). 
(d) Schematic plot of the time-averaged supercurrent density as a function of the longitudinal position. If the width of the stressed region is comparable to the charge imbalance equilibration length, then only one phase slip line will nucleate in the stressed region. 
}
\label{fig:hall004}
\end{figure}

For the reasons just exposed, it is clear that in plain NbSe$_2$ crystals the position where PSLs nucleate is sample-dependent and cannot be easily predicted. On the other hand, the mechanical flexibility of van der Waals SCs allows one to engineer mechanical stress in a narrow region across the whole sample width. Such region can act as a preferred nucleation site for a PSL. We chose to stamp the flake on a substrate with an etched trench, see Fig.~\ref{fig:hall004}(c). Van der Waals interaction with the substrate tends to pull the suspended part, which will result in tensile stress. If the width of the stressed region is smaller or comparable to $\Lambda_q$, we expect that only one PSL will nucleate there, see Fig.~\ref{fig:hall004}(d). New PSLs will then nucleate far from the former one, and at considerably higher current densities. 


Figure~\ref{fig:hall004} shows the results of four-terminal measurements on a trilayer NbSe$_2$ flake stamped on a substrate with prepatterned contacts and a 100~nm-wide etched trench in middle. Panel (a) shows the IVC for different temperatures, while panel (b) shows the color plot of the differential resistance as a function of current and temperature. For temperatures not too close to $T_c$, only one step in the IVC is visible. After this step the IVC is linear up to the experimentally accessible bias. Such linear regime extrapolates to a finite and large excess current. No other features are visible up to $I_{max}=300$~$\mu$A, the maximum current bias we applied. Only close to the critical temperature (for $T\geq 4.9$~K in Fig.~\ref{fig:hall004}(a)) it is possible to observe a second step within $I_{max}$, which indicates the nucleation of a second PSL.  After this second step, many PSLs precipitously nucleate and drive the sample into the normal state. 
The single isolated step is clearly evident in Fig.~\ref{fig:hall004}(b). The linear regime after the step is visible as a light blue region in the differential resistance color plot. This behavior is markedly different from that observed in plain devices, where several steps are visible immediately after the first (see Fig.~\ref{fig:4term}). Interestingly, the isolated PSL nucleating at the stressed line behaves as a dynamic Josephson junction~\cite{Sivakov2003} within the same crystal lattice. This scheme can therefore be useful to study the impact of the lattice orientation on the Josephson supercurrent, e.g.~by controlling the angle between the trench and the crystal axes. Moreover, it represents a fundamental building block for the implementation of more complex coherent devices within a single crystal as, e.g., a SQUID. 


In conclusion, we have investigated transport in few-layer NbSe$_2$ crystals under large current bias. Both two- and four-terminal measurements reveal that in this material phase slip lines are ubiquitous, as opposed to conventional metal films. Indeed, a \textit{direct} current-driven transition to the normal state is never observed.  In a submicron point contact the finite-bias conductance as a function of temperature can be successfully described with an adaptation of the  Skocpol-Beasley-Tinkham model. In extended flakes current-voltage characteristics also show strong signatures of phase slip line nucleation, whose details are, however, sample dependent. Finally, we demonstrate that mechanical strain can induce an artificial nucleation site for an isolated phase slip line.




\section*{Supporting Information Available}
PSL nucleation near the normal contact. Tunneling through the opaque contact \textit{6}. Parabolic behavior of the differential conductance at low bias. Supplementary data: four-terminal and two-terminal measurements. Temperature dependence of the critical current.

%

%
\begin{acknowledgments}
The work was funded by the Deutsche Forschungsgemeinschaft within Grants
DFG SFB1277 (B04) and GRK1570. Bulk NbSe$_2$ and hBN was purchased from HQ Graphene.
\end{acknowledgments}
\bibliographystyle{achemso}
\bibliography{biblio}
\end{document}